\documentclass[11pt]{amsart}

\usepackage{verbatim}
\usepackage{graphicx}
\numberwithin{equation}{section}
\newtheorem{theorem}{Theorem}[section]
\newtheorem{lemma}{Lemma}[section]

\newtheorem{definition}{Definition}[section]

\newtheorem{remark}{Remark}[section]

\newcommand{\dif}{{\mathrm d}}

\newcommand{\mspan}{{\mathrm span}}

\begin{document}
\date{\today}

\title[Traveling wave solutions for a predator-prey system]{Traveling wave solutions for a predator-prey system\\
 with Sigmoidal response function} 
 \thanks{Research of Lin was supported by the National Science Foundation under grant DMS-0708386;
Research of Weng was supported by the Natural Science Foundation of
China,  the research
fund for the Doctoral Program of Higher Education of China and NSF of Guangdong Province. }

\author{Xiaobiao Lin}
\address{ Department of Mathematics, North Carolina State University\\
 Raleigh, NC 27695-8205, USA}
\email{xblin@ncsu.edu(XL)  }

\author{Peixuan Weng}
\address{School of Mathematics, South China Normal University\\
Guangzhou 510631,  P.  R.  China}
\email{wengpx@scnu.edu.cn(PW)  }

\author{Chufen Wu}
\address{Department of Mathematics, Shanghai Jiaotong University\\
Shanghai 200030,  P.  R.  China}
\email{chufenwu@yahoo.cn(CW)   }



\begin{abstract}
We study the existence of traveling wave solutions for a diffusive
predator-prey system. The system considered in this paper is governed by a Sigmoidal response function which is more general than those
studied previously. Our method is an improvement to the original method introduced in the work of Dunbar \cite{Dunbar1,Dunbar2}. A bounded
Wazewski set is used in this work while unbounded Wazewski sets were
used in \cite{Dunbar1,Dunbar2}. The existence of traveling wave
solutions connecting two equilibria is established by using the
original Wazewski's theorem which is much simpler than the extended
version in Dunbar's work.

\end{abstract}

\maketitle

\section{Introduction}
\setcounter{equation}{0}\setcounter{theorem}{0}

Predator-prey models are important tools that help us to understand the bio and ecosystems surrounding us \cite{Hsu1}.
An important aspect of the model is how the predator interacts with the prey \cite{Fagan,Murray}, which can be described by
a functional response that specifies the rate of feeding of predator upon prey as a function of the prey density.

The existence of traveling wave solutions is an interesting feature in population dynamics and  plays an important
role in understanding the long time asymptotic properties of such systems. In his pioneering works \cite{Dunbar1, Dunbar2,Dunbar3}, Dunbar obtained the existence of several kinds of
traveling wave solutions for diffusive predator-prey systems with type I and type II functional responses. He
considered the existence of small
amplitude periodic traveling waves, and ``heteroclinic traveling waves'' that correspond to heteroclinic orbits
connecting equilibrium-to-equilibrium or equilibrium-to-periodic orbits. The methods used by Dunbar include the invariant manifold theory, the shooting method, Hopf bifurcation analysis, and LaSalle's invariance principle. Huang et al.
\cite{Huang} extended the work in \cite{Dunbar2} to $\mathbb{R}^4$  by using Dunbar's method in \cite{Dunbar3}. An interesting question is whether those results can be extended to a system with type III functional
response. Recently, Li et al. \cite{Li} proved the existence of
traveling waves in a diffusive predator-prey system with a
simplified type III functional response
$\varphi(u)=\frac{Bu^2}{1+Eu^2}$ by employing the method similar
to that used in \cite{Dunbar1, Dunbar2}.

The shooting method used by Dunbar is based on a variant of Wazewski's theorem
\cite{Dunbar1,Dunbar2,Dunbar3}. Similar to the original Wazewski set, Dunbar's  Wazewski set $\mathbb{W}$ has
the property that there is an orbit starting from  the unstable manifold of an equilibrium and staying in $\mathbb{W}$ in the future. However,
the Wazewski set $\mathbb{W}$ constructed in \cite{Dunbar1, Dunbar2,Dunbar3} was unbounded. In order to ensure the boundedness of the orbit, several
additional  Lemmas (see Lemma 7 in \cite{Dunbar1}, Lemma 10 in
\cite{Dunbar2}, Lemma 15 in \cite{Dunbar3}) were proved to rule out
the possibility that the constructed orbit may escape to infinity.
Although the topological idea of shooting method in \cite{Dunbar1,
Dunbar2,Dunbar3} is clear and elegant, the use of unbounded sets
$\mathbb{W}$ in $\mathbb{R}^3$ or $\mathbb{R}^4$ makes the argument long and hard to read.
Therefore, it is highly desirable to construct a simple, bounded Wazewski set
$\mathbb{W}$ and use the original Wazewski's theorem to simplify the proof of the existence of
heteroclinic traveling waves  related to various predator-prey systems.

 Another improvement is that in the proof of the existence of type I traveling waves,
we are able to treat the two cases $c> c_*$ and $c=c_*$ together, while in Dunbar's original work,
$c=c_*$ was treated by taking a sequence $c_n > c_*,\; c_n \to c_*$ and looking for convergence of the sequence of traveling wave solutions.
Finally, let $\gamma=\frac{e}{r}$, which is the quotient of the predator natural death rate
verses the prey growth rate. We have obtained an exact threshold value $\gamma^*$. If
$\gamma>\gamma^*$, the traveling waves are oscillatory. It means
if $\gamma$ is big, then the reason the limit of the wave exists is due to the
balance of large death rate of predator compensated by the
large growth rate of prey so the population can approach the interior
equilibrium. This causes the greater rate of exchange of
bio-mass between the predator and prey, therefore causes the
oscillation. While if $0<\gamma\leq \gamma^*$, the traveling waves are non-oscillatory. It means the
predators possess enough food to consume or have low death rate and then leads to the population density monotonically increasing.
In Dunbar's article, only a lower estimate for the threshold is given.

Consider the following diffusive predator-prey system
\begin{equation}\label{equ2}
\begin{aligned}
\frac{\partial N_1}{\partial t}&=D_1\frac{\partial^2 N_1}{\partial
x^2}+rN_1\left(1-\frac{N_1}{K}\right)-\frac{N_1^2}{a_1+b_1N_1+N_1^2}N_2\\
\frac{\partial N_2}{\partial t}&=D_2\frac{\partial^2 N_2}{\partial
x^2}+N_2\left(\frac{\alpha N_1^2}{a_1+b_1N_1+N_1^2}-e\right),
\end{aligned}
\end{equation}
where $N_1$ and $N_2$ are the population densities of the prey and  predator respectively. The functional response of the system
is more general and biologically useful than those in previous works.  To the best of our knowledge, no rigorous work has been done on the existence of traveling wave solutions for system \eqref{equ2}.

The shooting technique has been an important method in proving the existence of traveling waves solutions, e.g., see \cite{Dunbar1,Dunbar2,FL1, FL2, Huang,Li} and the references therein. The method used in this paper is motivated by the techniques used there.  However, as mentioned
above, we shall construct a bounded set $\mathbb{W}$ to replace the unbounded Wazewski set introduced in \cite{Dunbar1,Dunbar2,Huang,Li}. Assume that
our system has 3 equilibria $E$, $E_1$ and $E^*$. Instead of the extended
Wazewski's theorem used in \cite{Dunbar1,Dunbar2}, we shall use the original Wazewski's
topological principle as in \cite{Hartman} to prove the existence of an orbit connecting  $E_1$ and $E^*$  in $\mathbb{W}$. We first
 exam the flow on the surfaces of $\mathbb{W}$ to determine its egress sets. Then we
construct a curve $\mathcal{C}$ on the two dimensional unstable
manifold with two endpoints in two disjoint egress sets.
We show that there is a point on
$\mathcal{C}\cap\mathbb{W}$ such that the orbit started from this point
will remain in $\mathbb{W}$ by using Theorem 2.1 in Hartman
\cite[pp. 279]{Hartman}. A Liapunov function is constructed to prove that any orbit that remains in $\mathbb{W}$ shall approach an invariant set in $\mathbb{W}$ while the maximal invariant set in $\mathbb{W}$ is $E^*$.
Thus the solution that stays in $\mathbb{W}$ will approach the positive equilibrium $E^*$. In other words, it is a
heteroclinic orbit connecting the unstable equilibrium $E_1$ to $E^*$. We also show the existence of traveling wave solution connecting $E$ to $E^*$.  Our method is  more straightforward  than the shooting technique used in \cite{Dunbar1,Dunbar2,Huang,Li}.

In this paper we assume $D_1=0$. It is the limiting case where the prey species diffuses much slower than the predator species, e.g., a plant species being consumed by
a relatively mobile herbivore. The case $D_1$ is nonzero and small can be handle by the singular perturbation method and will appear in a separate paper (Wu et al. in preparation).

To write system  \eqref{equ2} in a
non-dimensional form, we rescale the variables
$$A=\frac{a_1}{K^2}, \; B=\frac{b_1}{K}, \;U=\frac{N_1}{K}, \;
W=\frac{ N_2}{rK}, \; t'=rt, \; x'=\sqrt{\frac{r}{D_2}}x.
$$
 By dropping the primes on
$t, x$ for notational convenience, \eqref{equ2} becomes
\begin{equation}\label{equ3}
\begin{aligned}
\frac{\partial U}{\partial t}&=U(1-U)-\frac{U^2W}{A+BU+U^2},\\
\frac{\partial W}{\partial t}&=\frac{\partial^2W}{\partial
x^2}+\frac{W}{r}\left(\frac{\alpha U^2}{A+BU+U^2}-e\right),
\end{aligned}
\end{equation}
where $r,e,\alpha>0, A\geq 0, B\geq 0$. Let $\beta=\frac{\alpha}{e}$. We require
that $\beta>1+A+B>1$, which ensures \eqref{equ3} has a positive
equilibrium corresponding to the coexistence of the two
species. It is easy to see that system \eqref{equ3} has three
spatially constant equilibria given by
\begin{equation}\label{equilibria}
\begin{aligned}
& E(0,0), \qquad  E_1(1,0) \qquad \mbox{and} \quad \quad
E^*(u^*,w^*), \\
\text{where }
& u^*=\frac{B+\sqrt{B^2+4A(\beta-1)}}{2(\beta-1)}, \quad
w^*=\frac{(1-u^*)[A+Bu^*+(u^*)^2]}{u^*}.
\end{aligned}
\end{equation}
By using the techniques
described above, we will establish the existence of traveling wave
solutions of system \eqref{equ3} connecting the equilibria $E_1$ and
$E^*$ (type I wave), which is called the ``waves of invasion" (see Chow and Tam
\cite{CT}, Shigesada and Kawasaki\cite{SK}).  This is of ecological interest since
it corresponds to a situation where an environment is initially inhabited
only by the prey species at its carrying capacity, the small invasion of
the predators drives the system to a new stable state of co-existence of both species.
We also establish the existence of traveling wave solutions of system \eqref{equ3}
connecting $E$ and $E^*$ (type II wave). This describes simultaneous spread of predator and prey into a new environment.
Moreover, a threshold property is given for
the two types of traveling waves to be oscillatory or non-oscillatory. See Theorem \ref{th2} for details.

The organization of the paper is as follows. In Section 2, we state
our main results on the existence of traveling wave solutions.
Section 3 and Section 4 are devoted to proving the existence or non-existence
for type I  and type II waves, respectively. The construction
of the Wazewski set $\mathbb{W}$ will also be given in Sections 3 and 4. In section 5, we discuss the threshold property for the oscillation
of the traveling waves in terms of $\gamma$, the quotient of the predator natural death rate
verses the prey growth rate.

\section{Main results}
\setcounter{equation}{0}\setcounter{theorem}{0}

A traveling wave solution of \eqref{equ3} is a solution of the
special form $U(t,x)=u(x+ct)=u(s)$ and $W(t,x)=w(x+ct)=w(s)$, where
$s=x+ct$ and $c>0$ is the wave speed. Substituting this
solution into \eqref{equ3}, we have the wave system
\begin{equation}\label{equ4}
\begin{aligned}
cu'(s)&=u(1-u)-\frac{u^2w}{A+Bu+u^2},\\
cw'(s)&=w''(s)+\frac{w}{r}\left(\frac{\alpha
u^2}{A+Bu+u^2}-e\right).
\end{aligned}
\end{equation}
Note that \eqref{equ4} also has three equilibria $E, E_1$ and $
E^*$, where $E$ corresponds to the absence of both species, $E_1$
corresponds to the prey at the environment carrying capacity in
the absence of the predator, and $E^*$ corresponds to the
co-existence of both species.
Rewrite system \eqref{equ4} as an equivalent system in $\mathbb{R}^3$
\begin{equation}\label{equ6}
\begin{aligned}
u'(s)&=\frac{1}{c}u\left(1-u-\frac{uw}{A+Bu+u^2}\right),\\
w'(s)&=z,\\
z'(s)&=cz+\gamma w\left(1-\frac{\beta u^2}{A+Bu+u^2}\right),
\end{aligned}
\end{equation}
where $\beta=\frac{\alpha}{e}, \gamma=\frac{e}{r}$. Then $E, E_1$ and $E^*$ correspond
to the critical points $(0,0,0), (1,0,0)$ and $(u^*,w^*,0)$ of
\eqref{equ6} respectively. In what follows, we shall still use  $E, E_1$ and $E^*$  to denote  $(0,0,0), (1,0,0)$ and $(u^*,w^*,0)$.
The Jacobian matrix of \eqref{equ6}
takes the form
\begin{equation}\label{1matrix}
\begin{pmatrix}
\frac{1}{c}(1-2u)-\frac{(2A+Bu)uw}{c(A+Bu+u^2)^2} & -\frac{u^2}{c(A+Bu+u^2)}  & 0 \\
0  & 0  & 1 \\
-\frac{\beta \gamma (2A+Bu)uw}{(A+Bu+u^2)^2} &
\gamma\left(1-\frac{\beta u^2}{A+Bu+u^2}\right) & c
\end{pmatrix}.
\end{equation}

We call the nonnegative solutions of \eqref{equ6} satisfying
\begin{equation}\label{equ6.1}
\lim\limits_{s\to-\infty}(u(s),w(s))=(1,0),\quad \lim\limits_{s\to
+\infty}(u(s),w(s))=(u^*,w^*)
\end{equation}
 type I waves, and the nonnegative solutions of \eqref{equ6}
satisfying
\begin{equation}\label{equ6.2}
\lim\limits_{s\to-\infty}(u(s),w(s))=(0,0),\quad \lim\limits_{s\to
+\infty}(u(s),w(s))=(u^*,w^*)
\end{equation}
 type II waves.

Consider the prey isocline of \eqref{equ4}
\[
w=h(u):=\frac{(1-u)(A+Bu+u^2)}{u}, \quad  u\in(0,+\infty).
\]
It is easy to check that $h'(u)=\frac{k(u)}{u^2}$, where
\[
k(u):=-2u^3+(1-B)u^2-A, \quad  u\in(0,+\infty).
\]
If $B\geq 1$, then $h(u)$ is monotone decreasing. Next, we consider the other case $0\leq B<1$. Obviously, there is a
negative real root to $k(u)=0$. Let $P:=(1-B)^2, Q:=-18A, R:=3A(1-B)$, then $\triangle:=Q^2-4PR=12A[27A-(1-B)^3]$.
(i) If $27A>(1-B)^3$, there is a complex conjugate
pair of roots with positive real part to $k(u)=0$. Hence, $k(u)<0$ for all $u\in(0,+\infty)$. (ii) If $27A=(1-B)^3$,
there is a double positive root $u=\frac{1-B}{3}$ to $k(u)=0$. It follows that $\frac{1-B}{3}$ is a local maximum point of $k(u)=0$
and thus $k(u)\leq 0$ for all $u\in(0,+\infty)$. (i) and (ii) imply that $h(u)$ is monotone decreasing. (iii) If $27A<(1-B)^3$,
there is three distinct real roots to $k(u)=0$.
Note that  $k'(u)=2u(1-B-3u)$,  then
$k(u)=0$ has two positive roots $\alpha_1, \alpha_2$ satisfying
$0<\alpha_1<\frac{1-B}{3}<\alpha_2<1$ (see Fig. 1). We now state the following result.

\begin{figure}[h]
\setlength{\unitlength}{1mm} \centering \centering
\begin{minipage}[h]{0.45\linewidth}
\centering
\begin{picture}(60,50)
\put(20,-5){\vector(0,1){45}} \put(0,17){\vector(1,0){60}}
\put(6,13){$\alpha_0$} \put(18,13){\footnotesize$0$}
\put(33,13){$\alpha_1$} \put(49,13){$\alpha_2$} \put(58,13){$u$}
\put(23,38){$w$} \qbezier(1,35)(10,-1)(33,17)
\qbezier(33,17)(50,33)(58,1)
\end{picture}
\caption{The graph of $k(u)$ for $0\leq B<1$ and $27A<(1-B)^3$}
\end{minipage}
\begin{minipage}[h]{0.04\linewidth}
\makebox[0.1cm]{}
\end{minipage}
\begin{minipage}[h]{0.45\linewidth}
\centering
\begin{picture}(65,50)
\put(8,-5){\vector(0,1){45}} \put(0,2){\vector(1,0){65}}
\put(6,-2){\footnotesize$0$} \put(10.5,-2){$u_0$}
\put(19,-2){$\alpha_1$} \put(42.5,-2){$\alpha_2$} \put(63,-2){$u$}
\put(57,-2){\footnotesize$1$} \put(11,38){$w$}
\qbezier(9,40)(10,2)(33,17) \qbezier(33,17)(50,32)(58,2)
\multiput(11.5,21.5)(0,-2){10}{\line(0,-1){1}}
\multiput(12,22)(2,0){16}{\line(1,0){1}}
\multiput(44,21.5)(0,-2){10}{\line(0,-1){1}}
\put(20.5,13){\line(0,-1){11}}
\multiput(20.5,12.6)(2.2,0){16}{\line(1,0){1}}
\multiput(54.2,12)(0,-1){10}{\line(0,-1){1}}\put(51.5,-2){$u_1$}
\end{picture}
\caption{The prey isocline $w=h(u)$ for $0<B<1$ and $27A<(1-B)^3$}
\end{minipage}
\end{figure}

\begin{lemma}\label{le1}
If $0\leq B<1$ and $27A<(1-B)^3$, then $k(u)=0$ has two positive roots
$\alpha_1, \alpha_2$ satisfying
$0<\alpha_1<\frac{1-B}{3}<\alpha_2<1$, and there exists a unique
$u_0\in(0,\alpha_1)$ satisfying
\begin{equation}\label{equ7}
h(u_0)=h(\alpha_2)
\end{equation}
or a unique
$u_1\in(\alpha_2,1)$ satisfying
\begin{equation}\label{equ7*}
h(u_1)=h(\alpha_1).
\end{equation}
\end{lemma}
\begin{proof}
Note that $h(1)=0$, $\lim_{u\to {0^+}}h(u)=+\infty$, $h'(u)>0$ for
$u\in(\alpha_1,\alpha_2)$ and $h'(u)<0$ for
$u\in(0,\alpha_1)\cup(\alpha_2,1)$. Here $\alpha_1 < \alpha_2$ are the roots for $h'(u)=\frac{k(u)}{u^2}=0$. Then $w=h(u)$ has a local minimum at  $\alpha_1$ and a local maximum at $\alpha_2$. Moreover, there exists a unique
$u_0\in(0,\alpha_1)$ satisfying $h(u_0)=h(\alpha_2)$ or a unique
$u_1\in(\alpha_2,1)$ satisfying $h(u_1)=h(\alpha_1)$  (see Fig. 2).
\end{proof}
\begin{lemma}\label{remark2.1}
Under any one of the following cases:\begin{enumerate}
\item[\rm{(i)}]  $B\geq 1$,
\item[\rm{(ii)}]  $0\leq B<1$, $27A\geq(1-B)^3$,
\item[\rm{(iii)}]  $0\leq B<1$,  $27A<(1-B)^3$, $u^*\leq u_0$ or $u^*\geq u_1$, where $u_0$ is given
by \eqref{equ7}, and $u_1$ by \eqref{equ7*},\end{enumerate}
 we have $(u-u^*)[h(u)-h(u^*)]\leq
0$ for $0<u<1$, where $(u^*,w^*)=(u^*,h(u^*))$ is the positive equilibrium of \eqref{equ4}.\end{lemma}

By computing the eigenvalues at $E_1,\;E$ and $E^*$ in sections 3, 4 and 5, we find that $E,$ $E_1,$ $E^*$ are all saddle points in the parameter ranges considered in this paper. The local unstable manifolds for $E$ and $E_1$ are two dimensional and the local stable manifold for $E^*$ is two dimensional. Generically, both type I and type II waves are transversal heteroclinic orbits. Therefore, we expect that both waves should exist for an open set of parameters, plus maybe some of its boundary points.

We now state our main results.
\begin{theorem}\label{th2}
Let $\beta>A+B+1>1$,  $c_*:=\sqrt{4\gamma\left(\frac{\beta}{A+B+1}-1\right)}$ and $(u^*, w^*)$ be as in \eqref{equilibria}.
\begin{enumerate}
\item If $0<c<c_*$,
then type II waves do exist while type I waves do not.

\item  (i) If $c\geq c_*$, $A, B$ satisfy one of the three conditions in Lemma \ref{remark2.1}, then type I waves do exit while type II waves do not.

(ii) If $c\geq c_*$, $0\leq B<1$,  $27A<(1-B)^3$ and $u_0< u^*<u_1$, then there is a traveling wave $\phi(s)$ with $\lim\limits_{s\to-\infty} \phi(s)=(1,0)$ and $\phi(s)\in \mathbb{W}_{(u,w)}=\{(u,w)|0\leq u\leq 1,  0\leq w\leq \beta(1+c^2/2)(1-u^*),\}$ for $s\geq 0$, while type II waves do not exist.

\item Let $(u,w)$ be a type I or type II  traveling wave. Then there exists a value $\gamma^*=\gamma^*(A,B,\beta,c)$ such that  if $0<\gamma\leq \gamma^*$, $(u,w)$ is non-oscillatory and approaches $(u^*,w^*)$
monotonically if $s$ is sufficiently large, while if $\gamma>\gamma^*$, $(u,w)$ have
exponentially damped oscillations about $(u^*,w^*)$ as $s\to\infty$.

Furthermore, we have
$$\gamma^*=\frac{c}{27\delta_2q(u^*)}[(2\omega_c^2+6\delta_1)\sqrt{\omega_c^2+3\delta_1}-(2\omega_c^3+9\delta_1\omega_c)],$$
where $\omega_c:=\frac{-\delta_1}{c}+c$, $\delta_1:=-k(u^*)q(u^*)$,  $\delta_2:=(2A+Bu^*)(1-u^*)$ and
$$ q(u):=\frac{1}{A+Bu+u^2},\quad k(u):=-2u^3+(1-B)u^2-A.$$

\end{enumerate}
\end{theorem}

\begin{remark} The results in \cite[Theorem 2.2]{Li} is a special case in (2) of Theorem \ref{th2}, i.e. $B=0$. Our contributions to this special case are as follows: (I) The existence of type I waves for cases $c=c_*$ and $c>c_*, u^*=u_1$ or $u^*\leq u_0$ was not in that paper. (II) Type II waves and the bounded waves in our case $B=0$,  $27A<(1-B)^3$, $u_0< u^*<u_1$ were not discussed in \cite{Li}. In fact, the latter cannot be obtained easily by the unbounded Wazewski set used in their work. (III) The non-monotone of traveling waves is not discussed in \cite{Li}.
\end{remark}

The proof of Theorem \ref{th2} will be given in the following three sections. In Section 3 and Section 4, we prove the assertion (1) and (2) respectively.   The procedure of proofs will be divided into several
steps with the main argument stated as several Lemmas for easy understanding.

Consider the differential equation
\begin{equation}\label{dif}
\textbf{y}'=\textbf{f}(t,\textbf{y}),
\end{equation}
where $\textbf{f}(t,\textbf{y})$ is a continuous function defined on
an open $(t,\textbf{y})$-set $\Omega$. Let $\Omega_0$ be an open
subset of $\Omega$, $\partial\Omega_0$ be the boundary and
$\bar{\Omega}_0$ be the closure of $\Omega_0$.

\begin{definition}\cite[pp. 278]{Hartman}
A point $(t_0,\textbf{y}_0)\in\Omega\cap\partial\Omega_0$ is called
an egress point of $\Omega_0$, with respect to \eqref{dif}, if for
every solution $\textbf{y}=\textbf{y}(t)$ of \eqref{dif} satisfying
\begin{equation}\label{dif1}
\textbf{y}(t_0)=\textbf{y}_0,
\end{equation}
there is an $\epsilon>0$ such that $(t,\textbf{y}(t))\in\Omega_0$
for $t_0-\epsilon\leq t<t_0$. An egress point  $(t_0,\textbf{y}_0)$
of $\Omega_0$ is called a strict egress point of $\Omega_0$ if
$(t,\textbf{y}(t))\not\in\bar{\Omega}_0$ for $t_0< t<t_0+\epsilon$
with a small $\epsilon>0$. The set of egress points of $\Omega_0$
will be denoted by $\Omega_0^e$ and the set of strict egress points
by $\Omega_0^{se}$. As a sufficient condition, the set of egress points
$\Omega_0^e$ can be determined by verifying
$\overrightarrow{f}\cdot\overrightarrow{n}>0$, where
$\overrightarrow{f}$ is the vector field, and $\overrightarrow{n}$ is
the outward normal vector of $\partial \Omega_0$.
\end{definition}

If $U$ is a topological space and $V$ a subset of $U$, a continuous
mapping $\pi:U\to V$ defined on all of $U$ is called a retraction of
$U$ onto $V$ if the restriction $\pi|_{V}$ of $V$ to $V$ is the
identity; i.e., $\pi(u)\in V$ for all $u\in U$ and $\pi(v)=v$ for
all $v\in V$. When there exists a retraction of $U$ onto $V$, $V$ is
called a retract of $U$.
\begin{lemma}\label{3.1}\cite[pp. 279]{Hartman}
Let $\textbf{f}(t,\textbf{y})$ be continuous on an open
$(t,\textbf{y})$-set $\Omega$ with the property that an initial value
determines a unique solution of \eqref{dif}. Let $\Omega_0$ be an open
subset of $\Omega$ satisfying $\Omega_0^e=\Omega_0^{se}$. Let $S$ be
a nonempty subset of $\Omega_0\cup\Omega_0^e$ such that
$S\cap\Omega_0^e$ is not a retract of $S$ but is a retract of
$\Omega_0^e$. Then there exists at least one point
$(t_0,\textbf{y}_0)\in S\cap\Omega_0$ such that the solution arc
$(t,\textbf{y}(t))$ of \eqref{dif}- \eqref{dif1} is contained in
$\Omega_0$ on its right maximal interval of existence.
\end{lemma}

\section{The existence of type I traveling waves}

At the equilibrium $(1,0,0)$, \eqref{1matrix} becomes
\begin{equation}\label{2matrix} J_{(1,0,0)}=
\begin{pmatrix}
-\frac{1}{c} & -\frac{1}{c(A+B+1)}  & 0 \\
0  & 0  & 1 \\
0 & \gamma\left(1-\frac{\beta }{A+B+1}\right) & c
\end{pmatrix}.
\end{equation}
Then by \eqref{2matrix}, we see that the eigenvalues of
 \eqref{equ6} at $(1,0,0)$ are
\[
\mu_1=-\frac{1}{c}, \quad
\mu_2=\frac{c-\sqrt{c^2-4\gamma\left(\frac{\beta}{A+B+1}-1\right)}}{2},
\quad
\mu_3=\frac{c+\sqrt{c^2-4\gamma\left(\frac{\beta}{A+B+1}-1\right)}}{2}.
\]
If $\mu_2 \neq \mu_3$, then the eigenvectors corresponding to $\mu_2, \mu_3$ are
\[
\textbf{X}_2=\left(\frac{-1}{A+B+1},1+c\mu_2,
\mu_2(1+c\mu_2)\right)^T,
\textbf{X}_3=\left(\frac{-1}{A+B+1},1+c\mu_3,
\mu_3(1+c\mu_3)\right)^T.
\]

\begin{proof}[
\textbf{Proof of the non-existence of the type I waves for
$0<c<c_*$}.].

If $0<c<c_*$, then $\mu_2$ and $\mu_3$ are complex. We shall show that the
heteroclinic orbit $\Gamma$ of \eqref{equ6} would oscillate around $w=0$
for large negative $s$.   The heteroclinic solution $\Gamma$
satisfies:
\begin{equation}\label{bvp}
w'(s)=z,\quad z'(s)=cz-\gamma w\left(\frac{\beta
u^2}{A+Bu+u^2}-1\right).
\end{equation}
 Let
$\ell:=\gamma(\frac{\beta
}{A+B+1}-1).$ Since $\beta>A+B+1$,  we know
 that $\ell>0$. In a small neighborhood of $(1,0,0)$, using the polar coordinates $w=\rho\sin(\theta), \; z=\rho\cos(\theta),$ we obtain
$$
\frac{\dif \theta}{\dif s}=\frac{z^2-cwz+\gamma w^2(\frac{\beta
u^2}{A+Bu+u^2}-1)}{{\rho}^2}.
$$
Note $c_*=\sqrt{4\ell}$. For any given $c\in (0,c_*)$, let $\varepsilon$ be a positive number such that $0<\varepsilon<\ell- \frac{c^2}{4}$. Since $\lim\limits_{s\to-\infty}u(s)=1$,
one can choose large $s_c>0$ such that
$$\ell-{\varepsilon}<\gamma(\frac{\beta
u^2(s)}{A+Bu(s)+u^2(s)}-1)<\ell+\varepsilon\mbox{ for }s<-s_c.$$
Therefore, for $s<-s_c$,  we have
\begin{equation}\label{theta}
\begin{aligned}
  \frac{\dif \theta}{\dif s}  >&\cos^2(\theta)-c\sin(\theta)\cos(\theta)+(\ell-{\varepsilon})\sin^2(\theta)\\
    =&[\cos(\theta)-\frac{c}{2}\sin(\theta)]^2+(\ell-\frac{c^2}{4} - \epsilon)\sin^2(\theta).\\
\end{aligned}
\end{equation}
The last expression is a periodic function of $\theta$ and is nonzero, which must be bounded below by a constant $c_0>0$. Therefore if $s<-s_c$, we have $\frac{\dif \theta}{\dif s} >c_0>0$, which can leads to
 $\lim\limits_{s\to-\infty}\theta(s)=-\infty$. From $w=\rho\sin(\theta)$, $w(s)$ would be negative for some $s<-s_c$. This violates the requirement that the traveling wave
solution considered should be nonnegative. The proof is complete.
\end{proof}

We now assume that $c\geq c_*$. Counting multiplicity, then there are three real
eigenvalues satisfying $\mu_1<0<\mu_2\leq \mu_3$.
If $c>c _*$, then $\mu_2 < \mu_3$, there are two eigenvectors $(\textbf{X}_2,\textbf{X}_3)$ corresponding to $(\mu_2, \mu_3)$. If $c=c_*$, then $\mu_2 = \mu_3$ is a double eigenvalue. There are two generalized eigenvectors $(\textbf{X}_2,\textbf{X}_3)$ corresponding to $\mu_2=\mu_3$.
By Theorems 6.1 and
6.2 in \cite[pp. 242-244]{Hartman}, there exists a two dimensional local
unstable manifold ${W}^{u}_{loc}(E_1)$ tangent to the span of
$\textbf{X}_2, \textbf{X}_3$.  The points on ${W}^{u}_{loc}(E_1)$ can be represented by the local coordinates $\Phi_2 : \mathbb{R}^2\rightarrow
\mathbb{R}^3$,
\[
\Phi_2(m,n)=(1,0,0)^T+m\cdot\textbf{X}_2+n\cdot\textbf{X}_3+o(|m|+|n|).
\]

Consider a prism shaped solid $\mathbb{W}$ (see Fig. 3) in
$(u,w,z)$ space bounded by the following five surfaces:
\begin{enumerate}
\item  The top surface $\mathbb{F}_t:=\{(u,w,z)\ |\ z=\frac{c}{2}w, 0<u<1, 0<w<w_m
\}$, where
$$w_m:=K^*(1-u^*),\quad
K^*:=\frac{(1+cd)[A+Bu^*+(u^*)^2]}{(u^*)^2}=(1+cd)\beta,\quad  d>\frac{c}{2}.$$
Namely, the quadrilateral $ABCFG$.
\item  The bottom surface $\mathbb{F}_b:=\{(u,w,z)\ |\ z=-\frac{\gamma}{c}w, 0<u<1, 0<w<w_m
\}$. Namely, the quadrilateral $ABCJH$.
\item  The front surface $\mathbb{F}_f:=\{(u,w,z)\ |\ w=K^*(1-u), u^*<u<1, -\frac{\gamma}{c}w<z<\frac{c}{2}w
\}$. Namely, the triangle $CFDJ$.
\item  The right vertical surface $\mathbb{F}_r:=\{(u,w,z)\ |\ w=w_m, 0<u<u^*,-\frac{\gamma}{c}w_m<z<\frac{c}{2}w_m
\}$. Namely, the quadrilateral $GFDJHE$, where one part $GFDE$ is
above the plane $z=0$, and the other part $EDJH$ is below the plane
$z=0$.
\item  The back surface $\mathbb{F}_k:=\{(u,w,z)\ |\ u=0, 0<w<w_m, -\frac{\gamma}{c}w<z<\frac{c}{2}w\}$. Namely, the triangle $AHG$.
\end{enumerate}

\begin{figure}[h]
\setlength{\unitlength}{1mm}
 \includegraphics[scale=0.43]{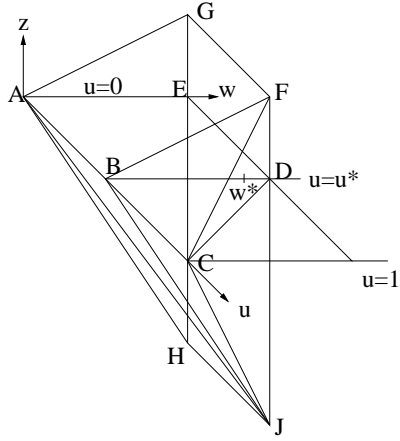}
 \caption{The graph of set $\mathbb{W}$ }\label{fig0}
\end{figure}

Note that $u^*\in(0,1)$, thus we obtain
$$w_m=K^*(1-u^*)=(1-u^*)\cdot \frac{(1+cd)[A+Bu^*+(u^*)^2]}{(u^*)^2}=\frac{1+cd}{u^*}w^*>w^*.$$
Next, we will prove the following lemma.

\begin{lemma} If an initial point $P_0=(u(0),w(0),z(0))$ is picking from
the interior of $\mathbb{W}$, then the flow $\phi(s,P_0) (s\geq 0)$ of  \eqref{equ6}
can only exit $\mathbb{W}$ from $\mathbb{F}_t$, $\mathbb{F}_b$,
$\mathbb{F}_r\cap\{z>0\}$ or the line segment $GF$.
\end{lemma}
\begin{proof}
(1) The outward normal vector of $\mathbb{F}_t$ is
$\overrightarrow{n_t}=(0,-\frac{c}{2},1)$, and the vector field is
$$\overrightarrow{f}=\left(\frac{u}{c}\left(1-u-\frac{uw}{A+Bu+u^2}\right),z,cz+\gamma w\left(1-\frac{\beta
u^2}{A+Bu+u^2}\right)\right).$$ Then it follows that
\begin{equation}\label{ww}\begin{array}{rl}
\overrightarrow{n_t}\cdot\overrightarrow{f}=&-\frac{c}{2}z+cz+\gamma w\left(1-\frac{\beta
u^2}{A+Bu+u^2}\right)\\
    =&\frac{c}{2}z+\gamma w\left(1-\frac{\beta
u^2}{A+Bu+u^2}\right)\\
  =&\frac{c^2}{4}w+\gamma w\left(1-\frac{\beta
u^2}{A+Bu+u^2}\right)\\
> &\frac{c^2}{4}w+\gamma w\left(1-\frac{\beta
}{A+B+1}\right)=w\left[\frac{c^2}{4}+\gamma \left(1-\frac{\beta
}{A+B+1}\right)\right] \geq 0
\end{array}\end{equation} by the assumption $ c\geq c_*$.
Therefore, the top surface  $\mathbb{F}_t$ belongs to the egress
set.

(2) The outward normal vector of $\mathbb{F}_b$ is
$\overrightarrow{n_b}=(0,-\frac{\gamma}{c},-1)$. Then it follows that
$$\begin{array}{rl}
\overrightarrow{n_b}\cdot\overrightarrow{f}=&-\frac{\gamma}{c}z-cz-\gamma w\left(1-\frac{\beta
u^2}{A+Bu+u^2}\right)\\
    =&\left(\frac{\gamma}{c}\right)^2w+\gamma w-\gamma w+\gamma w\frac{\beta
u^2}{A+Bu+u^2}\\
  =&\left(\frac{\gamma}{c}\right)^2w+\gamma w\frac{\beta
u^2}{A+Bu+u^2}>0,
\end{array}$$
and this implies that the bottom surface  $\mathbb{F}_b$ belongs to
the egress set.

(3) The outward normal vector of $\mathbb{F}_f$ is
$\overrightarrow{n_f}=(K^*,1,0)$. Then
$$\begin{array}{rl}
\overrightarrow{n_f}\cdot\overrightarrow{f}=&\frac{K^*}{c}u\left(1-u-\frac{uw}{A+Bu+u^2}\right)+z\\
    <&\frac{K^*}{c}u\left(1-u-\frac{uw}{A+Bu+u^2}\right)+\frac{c}{2}w\\
  =&\frac{K^*}{c}u\left[1-u-\frac{uK^*(1-u)}{A+Bu+u^2}\right]+\frac{c}{2}K^*(1-u)\\
  =& K^*(1-u)\left[\frac{u}{c}-\frac{K^*}{c}\frac{u^2}{A+Bu+u^2}+\frac{c}{2}\right]\\
  <& K^*(1-u)\left[\frac{1}{c}-\frac{K^*}{c}\frac{(u^*)^2}{A+Bu^*+(u^*)^2}+\frac{c}{2}\right]\\
 =& K^*(1-u)\left[\frac{1}{c}-\frac{1+cd}{c}+\frac{c}{2}\right]= K^*(1-u)\left(\frac{c}{2}-d\right)<0
\end{array}$$
by the assumption $d>\frac{c}{2}$.
Therefore, the  front surface  $\mathbb{F}_f$ belongs to the ingress set.

(4) The right vertical surface $\mathbb{F}_r$ is further divided into
$\mathbb{F}_r=\mathbb{F}_{r_1}\cup\mathbb{F}_{r_2}\cup\mathbb{F}_{r_3}$,
with $\mathbb{F}_{r_1}:=\mathbb{F}_r\cap\{z>0\}$, $\mathbb{F}_{r_2}:=\mathbb{F}_r\cap\{z<0\}$,
$\mathbb{F}_{r_3}:=\mathbb{F}_r\cap\{z=0\}$.
The outward normal vector of $\mathbb{F}_r$ is
$\overrightarrow{n_r}=(0,1,0)$. Then $\overrightarrow{n_r}\cdot\overrightarrow{f}=z$ and this implies that
$\mathbb{F}_{r_1}$ belongs to the egress set while $\mathbb{F}_{r_2}$ belongs to the ingress set.

(5) It is easy to see that the flow of \eqref{equ6} cannot enter or exit $\mathbb{W}$ through $\mathbb{F}_k$ since $u=0$ is invariant
 under the system \eqref{equ6}.

In order to check all the intersection part of two surfaces for the
solid $\mathbb{W}$, we make the following observation:

\emph{If $M$ and $N$ are two surfaces that intersect along a line
$\ell$, and if the vector field is transverse to these two surfaces,
then the line $\ell$ is on the egress set if and only if the vector
field points outward to both surfaces $M$ and $N$.}

From this observation, we see that the line segments  $GF$ is on the
egress set. It is easy to check that
$\mathbb{F}_{r_3}=\mathbb{F}_r\cap\{z=0\}$ is not on the egress set.
Furthermore, notice that $AC$ is a part of $u$ axis which is an
invariant set of \eqref{equ6}, which is not a part of the egress
set.

Thus, we have shown that $\mathbb{F}_t$, $\mathbb{F}_b$,
$\mathbb{F}_{r_1}=\mathbb{F}_r\cap\{z>0\}$ and the line segment $GF$
belong to the egress set, and no more. The proof of the lemma is
complete.
\end{proof}

\begin{definition} Consider a two-dimensional surface $\bar S=\text{span}\{\mathbf{u_1}, \mathbf{u_2}\}$ in $\mathbb{R}^3$  and let $\vec{n}$ be a normal vector to $\bar S$. Any vector $\mathbf{w}\notin \bar S$ is on the positive (or negative) side of $\bar S$ with respect to the normal $\vec{n}$ if $\vec{n}\cdot \mathbf{w} >0$ (or $<0$).

We say the two points $\mathbf{w_1}$ and $\mathbf{w_2}$ are on the same side of  $\bar S$ (or on the opposite side of $\bar S$) if the product $ (\vec{n}\cdot\mathbf{w_1})
(\vec{n} \cdot \mathbf{w_2})>0$ (or $<0$).
\end{definition}

In particular, if we choose $\vec{n}=\mathbf{u_1} \times \mathbf{u_2}$, then
$ \vec{n} \cdot \mathbf{w} = \det(\mathbf{u_1}, \mathbf{u_2}, \mathbf{w})$.

\begin{lemma}\label{le8}
There exists an arc  $\mathcal
{C}=\overline{C_1 C_2} \subset\mathbb{W}\cap {W}^{u}_{loc}(E_1)$ of which one end point
$C_1\in  \mathbb{F}_t$, and the other end point
$C_2\in  \mathbb{F}_b$.
\end{lemma}

\begin{proof} First, assume that $c>c_*$. Then $E_1$ has two distinct positive  eigenvalues. The two dimensional unstable manifold is tangent to
$S=\mspan(\textbf{X}_2,\textbf{X}_3)=\mspan(\textbf{X}_2,-\textbf{X}_3)$. Recall that
$\overrightarrow{CF}=(u^*-1,w_m,\frac{c}{2}w_m)^T$ and $
\overrightarrow{CJ}=(u^*-1,w_m,-\frac{\gamma}{c}w_m)^T$ are the vectors
corresponding to the rays $CF$ and $CJ$, and $\vec{u}=(1,0,0)^T$ be the
unit vector on the $u-$axis with a start point  $C=(1,0,0)$.

Note that $\beta>A+B+1, c>c_*$ and $d>\frac{c}{2}$, we obtain
$$\begin{array}{rl}
\det(\textbf{X}_2,\textbf{X}_3,\overrightarrow{CF})=&\frac{w_m}{A+B+1}\left(1+\frac{c^2}{2}\right)(\mu_3-\mu_2)-
(1-u^*)(1+c\mu_2)(1+c\mu_3)(\mu_3-\mu_2)\\
      =&(1-u^*)(\mu_3-\mu_2)\left[\frac{K^*}{A+B+1}\left(1+\frac{c^2}{2}\right)-(1+c\mu_2)(1+c\mu_3)\right]\\
  =&(1-u^*)(\mu_3-\mu_2)\left[\frac{(1+cd)\beta}{A+B+1}\left(1+\frac{c^2}{2}\right)-1-c^2-c^2\cdot\frac{c_*^2}{4}\right]\\
  >&(1-u^*)(\mu_3-\mu_2)\left[\left(1+c\cdot\frac{c}{2}\right)\left(1+\frac{c^2}{2}\right)-1-c^2-\frac{c^4}{4}\right]=0,
\end{array}$$
$\det(\textbf{X}_2,\textbf{X}_3,\vec{u})=(\mu_3-\mu_2)(1+c\mu_2)(1+c\mu_3)>0,$
$$\det(\textbf{X}_2,-\textbf{X}_3,\overrightarrow{CJ})=-(1-u^*)(\mu_3-\mu_2)
\left[\frac{K^*}{A+B+1}\left(1+\gamma+c^2\right)-(1+c\mu_2)(1+c\mu_3)\right]<0,$$
$\det(\textbf{X}_2,-\textbf{X}_3,\vec{u})=-\det(\textbf{X}_2,\textbf{X}_3,\vec{u})<0.$

This implies that $\vec{u}$ and $\overrightarrow{CF}$ are on the
same side of the tangent plane $S$. Also, $\vec{u}$ and
$\overrightarrow{CJ}$ are on the same side of $S$.

Next assume that $c=c_*$. Then $E_1$ has a double real eigenvalue $\mu_2 = \mu_3 = \frac{c}{2}$. The two dimensional
unstable manifold is tangent to
$S=\mspan(\textbf{X}_2,\textbf{X}_3)$, where $\textbf{X}_2$ and $\textbf{X}_3$ are generalized eigenvectors corresponding to $\mu_2$.
Denote the first row of the matrix
$$
 (\frac{c}{2} I -J_{(1,0,0)})^2 = \begin{pmatrix}\left(\frac{c}{2}+ \frac{1}{c}  \right)^2 &
\frac{ \frac{c}{2}+ \frac{1}{c}}{c(A+B+1)} + \frac{1}{2(A+B+1)} &\frac{-1}{c(A+B+1)}\\
0&0&0\\ 0&0&0 \end{pmatrix}.
$$
 by $\mathbf{r}_1$.  Since any generalized eigenvector $\textbf{X}$ satisfies $(\mu_2 I -J_{(1,0,0)})^2\textbf{X} = 0$, we see
 that $\mathbf{r}_1$ is a normal vector to $S$.
Then for any $\vec{v} \in \mathbb{R}^3$, whether $\vec{v}$ is on which side of $S$ can be determined by
the sign of the inner product $\mathbf{r}_1 \cdot \vec{v}$. Recall that $\vec{u}=(1,0,0)^T$ and
$\frac{\overrightarrow{CF}}{w_m}=(-\frac{1}{K^*},1, \frac{c}{2})^T$, it is easily checked
$$\mathbf{r}_1\cdot\vec{u} = \left(\frac{c}{2}+ \frac{1}{c}
\right)^2>0,
$$
\begin{align*}
&\mathbf{r}_1 \cdot\frac{ \overrightarrow{CF}}{w_m}\\
=&\left[\frac{ \frac{c}{2}+ \frac{1}{c}}{c(A+B+1)} + \frac{1}{2(A+B+1)}   \right] - \frac{(\frac{c}{2}+\frac{1}{c})^2 }
{K^* } -\frac{1}{2(A+B+1)}\\
=&(\frac{c}{2}+\frac{1}{c}) \cdot \frac{K^* - (A+B+1)(1+\frac{c^2}{2})}{c(A+B+1)K^*}.
\end{align*}
Since $K^* = (1+cd)(\frac{A}{(u^*)^2} + \frac{B}{u^*}+1)>(1+cd)(A+B+1)\geq (1+\frac{c^2}{2})(A+B+1)$,
we have $\mathbf{r}_1 \cdot \overrightarrow{CF} >0$.

Next recall that $\frac{\overrightarrow{CJ}}{w_m} = (-\frac{1}{K^*},1,-\frac{\gamma}{c})^T$. It is easily verified
\begin{align*}
\mathbf{r}_1 \cdot \frac{\overrightarrow{CJ}}{w_m} > \mathbf{r}_1 \cdot\frac{ \overrightarrow{CF}}{w_m}>0.
\end{align*}

In conclusion, $\vec{u}$, $\overrightarrow{CF}$ and
$\overrightarrow{CJ}$ are on the same side of $S$ just like the case $c>c_*$.
Denote this side as  the ``positive-$u$ side of $S$''.

In geometry, same notations are often used for vectors and rays. In the rest of the section, we assume
$(\overrightarrow{CF},\overrightarrow{CJ},\vec{u})$ are three rays with vertex $C$. Let
co$(\overrightarrow{CF},\overrightarrow{CJ},\vec{u})$ be the convex hull generated by
the rays $\overrightarrow{CF}$, $\overrightarrow{CJ}$ and the ray pointing to the positive $u-$axis with the
starting point $C$. This is an infinite cone of triangular cross section and with the vertex $C$.
 Then except for the point $C$,
co$(\overrightarrow{CF},\overrightarrow{CJ},\vec{u})$  is on the ``positive-$u$ side'' of the tangent plane $S$.

Define
\begin{align*}
S_1:=\{(u,w,z)~|~z=\frac{c}{2}w \},\quad
S_2:=\{(u,w,z)~|~z=-\frac{\gamma}{c}w\}.
\end{align*}
Note that $\mathbb{F}_t$ ($\mathbb{F}_b$) is a part of $S_1$ ($S_2$).

Note the plane $S$
intersects with $S_1$ transversely and $S$ is the
tangent plane of $W^u_{loc}(E_1)$ at $E_1$, the unstable manifold
$W^u_{loc}(E_1)$ intersects with $S_1$ on a smooth line segment $\ell_1$. If
we make $\ell_1$ sufficiently short, then $\ell_1$ does not enter
$co(\overrightarrow{CF},\overrightarrow{CJ},\vec{u})$. Denote by $\ell_t$ the part of $\ell_1$ where
$z>0$ and which is on $\mathbb{F}_t$. We now select
$C_1$ on $\ell_t$. It is clear that $C_1\in W^u_{loc}(E_1)\cap
\mathbb{F}_t$ but not in $co(\overrightarrow{CF},\overrightarrow{CJ},\vec{u})$.

Similarly, by using the property of $S_2$, we can construct a short
line segment $\ell_b$ which is on $W^u_{loc}(E_1)\cap \mathbb{F}_b$
but not in $co(\overrightarrow{CF},\overrightarrow{CJ},\vec{u})$. Select a point $C_2\in \ell_b$.
Using the local coordinates $u=\tilde u(w,z)$ for $W^u_{loc}(E_1)$, we can construct a curve segment $\mathcal{C}=\overline{C_1 C_2}$ on  $W^u_{loc}(E_1)$, connecting $C_1$ and $C_2$ and is between the two surfaces $\mathbb{F}_t$ and $\mathbb{F}_b$.

Let $d(P_1,P_2)$ be the distance function  in $\mathbb{R}^3$. Define the minimum distance
$$ \eta:= \min_{P_1,P_2}\{d(P_1,P_2)\;|\;\overrightarrow{CP_1}\in co(\overrightarrow{CF},
 \overrightarrow{CJ},\vec{u}), \|CP_1\| = 1, P_2 \in S\}.
$$
 Then $0<\eta\leq 1$. If $\overrightarrow{CQ}$ is a nonzero vector in $co(\overrightarrow{CF},
 \overrightarrow{CJ},\vec{u})$,   then the distance $d(Q,S) \geq \eta \rho$, where $\rho=\|{CQ}\|$.
  On the other hand, since  $W^u_{loc}(E_1)$ is tangent to $S$, the distance of any point $Q\in \overline{C_1 C_2}$
   to $S$ is $O(\rho^2)$. If $\rho$ is sufficiently small, $O(\rho^2) < \eta \rho$.
   Based on this, we find that the entire  curve $\overline{C_1 C_2}$ is inside $\mathbb{W}$ if its distance
   to the vertex $C$ is sufficiently small.
 \end{proof}

\begin{lemma}\label{le9}
There is a point $P_0=(u_0,w_0,z_0)\in \mathcal {C}\cap\mathbb{W}$
such that the flow $\phi(s,P_0)$ will remain in $\mathbb{W}$ for all
$s\geq 0$.
\end{lemma}
\begin{proof} Note that $A+Bu+u^2=0$ has roots
$\frac{-B\pm\sqrt{B^2-4A}}{2}$. Set
\[
\hat{u}=\begin{cases}
\frac{-B+\sqrt{B^2-4A}}{2}, & B^2-4A\geq 0,\\
-\infty, & B^2-4A< 0.
\end{cases}
\]
Define $\mathcal{D}=(\hat{u},\infty)$, then $u',z'$ has no singularity
on $\mathcal{D}$. Let $\Omega=\mathcal{D}\times\mathbb{R}^2$ and
$\Omega_0=\mathbb{W}$. Then $\Omega_0^e=\Omega_0^{se}=\mathbb{F}_t\cup(\mathbb{F}_{r}\cap\{z>0\})\cup GF\cup\mathbb{F}_b$.
Also, $C_1\cup C_2$ is not a retract of
$\mathcal {C}$ but a retract of $\Omega_0^e$. By Lemma \ref{3.1},
there is a point $P_0\in \mathcal {C}\cap\mathbb{W}$ such that the
flow $\phi(s,P_0)$ will remain in $\mathbb{W}$ for all $s\geq 0$.
\end{proof}

\begin{lemma}\label{le11}
Let $P_0$ be defined as in Lemma \ref{le9}. Then  $\phi(s,P_0)\rightarrow (u^*,w^*,0)^T$
as $s\rightarrow +\infty$.
\end{lemma}
\begin{proof}
Let $u^-$ be the negative root of the equation $1-\frac{\beta u^2}{A+Bu + u^2} = 0$, then the two roots of  $1-\frac{\beta u^2}{A+Bu + u^2} = 0$ are $u^*$ and $u^-$.  Define a Liapunov function
\[
V(u,w,z)=c\gamma\int_{u^*}^u
\frac{(\xi-u^*)(\xi-u^-)}{\xi^2}\dif
\xi+[cw-w^*-z]+w^*\left[\frac{z}{w}-c\log\frac{w}{w^*}\right].
\]
It is easy to check that $V(u,w,z)$ is continuously differentiable and bounded below on the compact set $\mathbb{W}$. Moreover,
$$\begin{array}{rl}
\frac{\dif V}{\dif s}=&\frac{\partial V}{\partial u}u'+\frac{\partial V}{\partial w}w'
+\frac{\partial V}{\partial z}z'\\
    =&\frac{c\gamma(u-u^*)(u-u^-)}{u^2}\cdot
    \frac{u}{c}\left(1-u-\frac{uw}{A+Bu+u^2}\right)+\left(c-\frac{w^*z}{w^2}-\frac{cw^*}{w}\right)z\\
    &\ +\left(\frac{w^*}{w}-1\right)\left[cz+\gamma w\left(1-\frac{\beta u^2}{A+Bu+u^2}\right)\right]\\
  =&\frac{\gamma(u-u^*)(u-u^-)}{A+Bu+u^2}[h(u)-w]+\gamma\left[\frac{\beta u^2}{A+Bu+u^2}-1\right]
  (w-w^*)-\frac{w^*z^2}{w^2}\\
  =&\frac{\gamma(u-u^*)(u-u^-)}{A+Bu+u^2}[h(u)-w]+\frac{\gamma(u-u^*)(u-u^-)}
  {A+Bu+u^2}(w-w^*)-\frac{w^*z^2}{w^2}\\
  =&\frac{\gamma(u-u^*)(u-u^-)}{A+Bu+u^2}[h(u)-w^*]-\frac{w^*z^2}{w^2}\\
  =&\frac{\gamma(u-u^-)}{A+Bu+u^2}(u-u^*)[h(u)-w^*]-\frac{w^*z^2}{w^2},\\
\end{array}$$
where $h(u)$ is defined in Section 2 and $w^*=h(u^*)$. By Lemma \ref{remark2.1}, it immediately follows that  $(u-u^*)[h(u)-w^*]\leq
0$ for $0<u<1$. Furthermore, $\frac{\dif V}{\dif s}=0$ if
and only if $\{u=u^*, 0<w< w_m, z=0\}$. The largest invariant subset of
this line segment in $\mathbb{W}$ is the positive equilibrium
$(u^*,w^*,0)$. By the LaSalle$^{'}$s invariance principle
\cite{LaSalle}, it follows that $\phi(s,P_0)\rightarrow
(u^*,w^*,0)^T$ as $s\rightarrow +\infty$.
\end{proof}

\begin{proof}[\textbf{Proof of existence of traveling waves for $c\geq c_*$.}]

Assume that  one of the conditions in Lemma \ref{remark2.1} holds. Choose a point $P_0$ in $\mathcal {C}\cap \mathbb{W}$ as defined by
Lemma \ref{le9}. By Lemmas \ref{le9}--\ref{le11}, we see that
$\phi(s,P_0)$ will remain in $\mathbb{W}$ and further approach the
positive equilibrium $(u^*,w^*,0)^T$. Also, $\phi(s,P_0)\rightarrow
(1,0,0)^T$ as $s\to-\infty$ since $P_0\in W^u_{loc}(E_1)$. Thus, a type I traveling
wave solution has been constructed for $c\geq c_*$.

If $0\leq B<1$,  $27A<(1-B)^3$, $u_0< u^*<u_1$, then by Lemma \ref{le9}, there is a traveling wave $\phi(s,P_0)$ with $\lim\limits_{s\to-\infty} \phi(s,P_0)=(1,0,0)$ and $\phi(s,P_0)\in \mathbb{W}$  for $s\geq 0$. Note that the projection of $\mathbb{W}$ on $(u,w)$ plane is in the rectangle $\mathbb{W}_{(u,w)}:=\{(u,w)|0\leq u\leq 1,  0\leq w\leq K^*(1-u^*),\}$, where $K^*=\beta(1+cd)$, $d$ is any constant larger than $c/2$.
The proof has been completed.
\end{proof}

\section{The existence of type II traveling waves}

At the equilibrium $E=(0,0,0)$, \eqref{1matrix} becomes
\begin{equation}\label{3matrix}
J_{(0,0,0)}=
\begin{pmatrix}
\frac{1}{c} & 0  & 0 \\
0  & 0  & 1 \\
0 & \gamma & c
\end{pmatrix}.
\end{equation}
Then the eigenvalues of \eqref{equ6} at $E$ are
\[
\nu_1=\frac{c-\sqrt{c^2+4\gamma}}{2}<0, \quad
\nu_2=\frac{1}{c}>0, \quad
\nu_3=\frac{c+\sqrt{c^2+4\gamma}}{2}>0.
\]
Thus there is a two dimensional local unstable manifold ${W}^u_{loc}(E)$
based at $E$. The eigenvectors corresponding to $\nu_2, \nu_3$
are respectively $\textbf{Y}_2=(1,0,0)^T, \textbf{Y}_3=(0,1,
\nu_3)^T$. Let $L=(0,1,0)^T$ be a vector that is complementary to the plane $S=\text{span}\{\textbf{Y}_2,\textbf{Y}_3\}$.
In a small neighborhood of $E$, the points on ${W}^{u}_{loc}(E)$ can be expressed as:
$$
(u,w,z)^T=(0,0,0)^T+m\cdot\textbf{Y}_2+n\cdot\textbf{Y}_3+\ell^*(m,n) L,
$$
where $\ell^*(m,n)=O(m^2 + n^2)$ is a smooth function of $(m,n)$.
Since the $u$-axis is invariant under the flow
 of \eqref{equ6}, if $m$ is small, then $m \textbf{Y}_2\in W^u_{loc}(E)$, that is, $\ell^*(m,0)=0$.

On the other hand, equation \eqref{equ6} is linear if $u=0$. For any $n\in \mathbb{R}$,  the line $n \textbf{Y}_3$ is
 invariant under the flow and $n \textbf{Y}_3 \in W^u_{loc}(E)$.  This shows $\ell^*(0,n) =0$.  Based on
 $\ell^*(m,0) =\ell^*(0,n) =0$, we have a better  estimate $ \ell^*(m,n) =  O(|mn|)$. And any point on $W^u_{loc}(E)$
 can be expressed as
\begin{equation}\label{mn1}
Q(m,n)=(u,w,z)^T=(0,0,0)^T+m\cdot\textbf{Y}_2+n\cdot\textbf{Y}_3+(O(|mn|)) \cdot L.
\end{equation}

\begin{proof}[\textbf{Proof of the non-existence of the type II waves for $c\geq c_*$.}]

  We present an indirect proof. If there is a heteroclinic solution $\{\textbf{y}(s)=(u(s),w(s),z(s))^T: s\in \mathbb{R}\}$ connecting $E$ to $E^*$, then  for  sufficiently large negative $s_1$, we have $ \textbf{y}(s_1) \in W^u_{loc}(E)$. Let $(m(s), n(s))$ be the parameter representation for $\mathbf{y}(s)$ as in \eqref{mn1}. Then we claim that $n(s_1) >0$. For from \eqref{mn1}, if $n(s_1)<0$, then $w(s_1)<0$, which is meaningless in the biological context. If $n(s_1)=0$, then  $w(s_1) = z(s_1)=0$, thus $\mathbf{y}(s_1)$ is on the $u$-axis. However, the $u$-axis is an invariant manifold connecting $E$ to $E_1$, this is a contradiction.

Using $n(s_1)>0$,  we have $z(s_1)=\nu_3
w(s_1)+o(|n|)>\frac{c}{2}w(s_1)$. We can show that
$z(s)>\frac{c}{2}w(s)$ and $w(s)>0$ for all $s>s_1$.  Here is the
proof.

Consider the open set $\Lambda:=\{0<u<1, w > 0,  z > \frac{c}{2}w\}$. Let $\chi(s) = z(s) - \frac{c}{2} w(s)$. From \eqref{equ6}, we calculate $\chi'(s)$ in $\Lambda$:
\begin{align*} \chi'(s)=
cz(s)+\gamma w(s)\left[1-\frac{\beta
u^2(s)}{A+Bu(s)+u^2(s)}\right]-\frac{c}{2}z(s)\\
> w(s)\left[\frac{c^2}{4}+\gamma\left(1-\frac{\beta
u^2(s)}{A+Bu(s)+u^2(s)}\right)\right]\\
\geq w(s)\left[\frac{c_*^2}{4}+\gamma\left(1-\frac{\beta}{A+B+1}\right)\right]=0.
\end{align*}

Now the heteroclinic solution $\mathbf{y}(s)$ satisfies  $\mathbf{y}(s_1) \in\Lambda$ and $0<u(s) < 1$ for all $s$. We can prove $\mathbf{y}(s) \in\Lambda$ for all $s > s_1$ by contradiction.  Assume that $s_2>s_1$ is the first time that $\mathbf{y}(s)$ hits the boundary of $\Lambda$. Then either
(1) $w(s_2) = 0$ or (2) $\chi(s_2) = 0$. Case (1) is impossible since for $s_1 < s < s_2$, $w'(s) = z(s) >0 $ in $\Lambda$. Case (2) is also impossible since for $s_1 < s < s_2$, $\chi'(s) >0$ which leads to $\chi(s_2) > \chi(s_1) > 0$.

However, if $\mathbf{y}(s) \in\Lambda$ for $s \geq s_1$, then $\lim_{s\to\infty} \mathbf{y}(s) \neq E^*$. The proof has been completed.
\end{proof}

\textbf{The existence of  the type II waves for $0<c< c_*$.}
 Define a solid $\mathbb{W}_0$ which is a modification of $\mathbb{W}$ as in \S 3, Fig 3. Compared to $\mathbb{W}$, the top surface is replaced by $\mathbb{F}_t:=\{(u,w,z)\ |\  0<u<1, 0<w<w_m, z=dw\}$ where $d>\nu_3$; the right vertical surface is replaced by  $\mathbb{F}_r:=\{(u,w,z)\ |\ 0<u<u^*, w=w_m,  -\frac{\gamma}{c} w_m < z < d w_m\}$; and the front surface is replaced by $\mathbb{F}_f:=\{(u,w,z)\ |\ u^*<u<1, w=K^*(1-u),  -\frac{\gamma}{c}w<z<d w
\}$. The back surface is $\mathbb{F}_k:=\{(u,w,z)\ |\ u=0, 0<w<w_m, -\frac{\gamma}{c}w<z< d w\}$.

The bottom surface is unchanged.

\begin{lemma}\label{le15}
 The egress sets for $\mathbb{W}_0$ are the bottom surface $\mathbb{F}_b$  and part of the right vertical surface $\mathbb{F}_r \cap \{z>0\}$.

\end{lemma}

\begin{proof}  Notice that $\{u=0\}$ is an
invariant set thus the back surface $\mathbb{F}_k$ is not an egress set.

The set $\mathbb{F}_r\cap \{z>0\}$ is obviously an egress set due to $w'=z>0$ there.

The bottom surface $\mathbb{F}_b$ remains the same and hence is an egress set.

The front surface $\mathbb{F}_f$ is larger than that of $\mathbb{W}$. However its normal vector remains the same. Modifying the estimates in the proof of Lemma 3.1, we have
$$\begin{array}{rl}
\overrightarrow{n_f}\cdot\overrightarrow{f}=&\frac{K^*}{c}u\left(1-u-\frac{uw}{A+Bu+u^2}\right)+z\\
    <&\frac{K^*}{c}u\left(1-u-\frac{uw}{A+Bu+u^2}\right)+d w\\
  =&\frac{K^*}{c}u\left[1-u-\frac{uK^*(1-u)}{A+Bu+u^2}\right]+ d K^*(1-u)\\
  <& K^*(1-u)\left[\frac{1}{c}-\frac{1+cd}{c}+d\right]= 0.
\end{array}$$
Thus, $\mathbb{F}_f$ is still an ingress set.

Since $\nu_3<d$, similar to the proof of Lemma 3.1, we have
\begin{align*}
\overrightarrow{n_t}\cdot\overrightarrow{f}=&-dz+cz+\gamma w\left(1-\frac{\beta
u^2}{A+Bu+u^2}\right)\\
    =&(-d+c)dw+\gamma w\left(1-\frac{\beta
u^2}{A+Bu+u^2}\right)\\
  =&w\left[-d^2+cd+\gamma\left(1-\frac{\beta
u^2}{A+Bu+u^2}\right)\right]\\
< &-w(d^2-cd-\gamma)\\
<&-w(\nu_3^2-c\nu_3 -\gamma) =0.
\end{align*}
Hence, $\mathbb{F}_t$ belongs to the ingress set.

Whether the  edges of $\mathbb{W}_0$ is an egress set can be checked easily. In particular, the line $GF$ as in Fig. 3 is not an egress set.

\end{proof}

In order to use Lemmma \ref{3.1}, we shall construct a curve $\mathcal {E} \subset W^u(E)\cap \mathbb{W}_0$, of which the two end points belong to the two disjoint egress sets  of $\mathbb{W}_0$.

\begin{lemma}\label{le14}
Assume that $0<c<c_*$, then there exists an
$P_1\in{W}^u_{loc}(E)\cap\Theta$ such that the flow $\phi(s,P_1)$
enters $\mathcal{Q}=\{(u,w,z)| u>u^*, w<w^*, z<0\}$ for some finite
$s=\bar{s}$, where $\Theta=\{(u,w,z)| u\geq 0, w\geq
0, z\geq 0\}$.
\end{lemma}
\begin{proof}
The proof is similar to that Dunbar in Lemma 10 \cite{Dunbar1}, and
shall be skipped.
\end{proof}

Choose a constant $\bar w$ such that $w(\bar{s}) < \bar w < w^*$. Define a solid as $\mathbb{W}_2:=co(B,C,D,J)\cap \{w<\bar w\}$.

\begin{lemma}\label{le16}
The only egress set for $\mathbb{W}_2$ is the egress set  $\mathbb{F}_b$ for $\mathbb{W}_0$.
\end{lemma}
\begin{proof}
The surface $w=\bar w, z<0$, where $w'<0$, cannot be an egress set for $\mathbb{W}_2$.

The surface $CDJ$ is part of $\mathbb{F}_f$ for $\mathbb{W}_0$ thus it cannot be an egress for $\mathbb{W}_2$.

The surface $BDJ \cap \{w<\bar w\}$ cannot be an egress set since $u'=0$ at $(u=u^*,w=w^*)$ from the first equation of \eqref{equ6}. We now have $u=u^*$ but $w<w^*$, therefore $u'>0$ there.

On the surface $BCD\cap \{w<\bar w\},\; z=0$. From \eqref{equ6}, $z'(s) =\gamma w\left(1-\frac{\beta u^2}{A+Bu+u^2}\right)<0$, since it would be $0$ if $u=u^*$. But now we have $u>u^*$.

\end{proof}

\begin{proof}[
\textbf{Proof of the existence of the type II waves for $0<c< c_*$.}]
Let $P_1$ be the point as in Lemma \ref{le14}. The orbit $\phi(s, P_1)$ enters $\mathbb{W}_0$ at some $s=s_3\geq \bar{s}$. Let $P_3 = \phi(s_3,P_1)$. The flow $\phi(s,P_3)$ cannot stay in $\mathbb{W}_2$ for all $s>s_3$ because $w'(s)<0,z'(s)<0$ there. Therefore it must exit through its egress set $\mathbb{F}_b$ as from lemma \ref{le16}.

On the other hand, from \eqref{mn1}, for a small real $n>0, \; Q(0,n):=n \textbf{Y}_3$ is on $W^u_{loc}(E)$. Moreover, from \eqref{equ6}, the solution $\phi(s, Q(0,n))$ hits the surface $w=w_m$  transversely in finite time. From the expression \eqref{mn1}, there is a small $m>0$ such that the corresponding $Q(m,n)$ is near $Q(0,n)$ and its $u$-coordinate is small and positive. Thus, $Q(m,n)\in W^u_{loc}(E)\cap \mathbb{W}_0$ and the flow $\phi(s,Q(m,n))$ hits the surface $w=w_m$ in finite time and its $u$-coordinate is small and positive. Let $P_2 = Q(m,n)$. Then  $\phi(s,P_2)$ exits $\mathbb{W}_0$ at $\mathbb{F}_r\cap \{z>0\}$.

With the given end points $P_1$ and $P_2$ and using the local coordinates for $W^u_{loc}(E)$, we can construct a small curve $\overline{P_1P_2}\subset W^u_{loc}(E)\cap \mathbb{W}_0$. Let $s_4$ be the first time that $\phi(s,P_1)$ hits $\mathbb{F}_b$ and let $s_5$ be the first time that $\phi(s,P_2)$ hits $\mathbb{F}_r\cap \{z>0\}$. Then
the curve $\mathcal {E}$ can be obtained as the union of three curves
$$\mathcal {E}:= \{\phi(s,P_1): 0 \leq s \leq s_4\} \cup \overline{P_1P_2} \cup
\{\phi(s,P_2): 0 \leq s \leq s_5\}.
$$
 From Lemma \ref{3.1}, there exits a point $P_0 \in \overline{P_1 P_2}$ such that the flow $\phi(s,P_0)$ remains in $\mathbb{W}_0$ for $s>0$. By using the same Liapunov function as in Lemma \ref{le11}, we see that $\phi(s,P_0) \to E^*$ as $s\to\infty$. The proof has been completed.

\end{proof}

\section{Oscillation of the traveling wave solutions}

Evaluating the Jacobian matrix of \eqref{equ6} at $E^*=(u^*,w^*,0)$, we have
\begin{equation}\label{matrix}
J_{(u^*,w^*,0)}=
\begin{pmatrix}
\frac{1}{c}k(u^*)q(u^*) & -\frac{1}{c}(u^*)^2q(u^*)  & 0 \\
0  & 0  & 1 \\
-\beta \gamma (2A+Bu^*)(1-u^*)q(u^*) & 0 & c
\end{pmatrix},
\end{equation}
where $k(u)=-2u^3+(1-B)u^2-A$ (see Lemma \ref{le1}) and
$q(u)=\frac{1}{A+Bu+u^2}$. Observe that $\beta (u^*)^2q(u^*) =1$. Using $\gamma$ as a parameter, the characteristic polynomial of
$J_{(u^*,w^*,0)}$ is
\[
p(\lambda,\gamma)=\lambda\left[\lambda-\frac{k(u^*)q(u^*)}{c}\right](c-\lambda)+\frac{\gamma}{c}(2A+Bu^*)(1-u^*)q(u^*).
\]
Note that if $B\geq 1$, then $k(u^*)<0$. If $0\leq B<1, \triangle\geq 0$, then $k(u^*)<0$ also holds. 
If $0\leq B<1, \triangle<0$, $0<u^*\leq u_0<\alpha_1$ and
$\alpha_2<u_1\leq u^*<1$, we have $k(u^*)<0$ (see page 6 for details)
and thus $-k(u^*)>0$. Let $\delta_1:=-k(u^*)q(u^*)$ and
$\delta_2:=(2A+Bu^*)(1-u^*)$. Then $\delta_1>0$, $\delta_2>0$
and
\[
p(\lambda,\gamma)=-\lambda^3+\left(\frac{-\delta_1}{c}+c\right)\lambda^2+\delta_1\lambda+\frac{\gamma}{c}\delta_2q(u^*).
\]

\begin{proof}[
\textbf{Proof of (3) in Theorem \ref{th2}}.]
Based on
\begin{equation}\label{derivative}
 p'(\lambda,\gamma) = -3 \lambda^2 +2\left(\frac{-\delta_1}{c}+c\right)\lambda +\delta_1,
\end{equation}
we find that $p'(0,\gamma) >0$ and $p'(\lambda,\gamma)$ has two critical points $\lambda_-< 0 < \lambda_+$.
 Together with the fact that on the real line, $p(\lambda,\gamma) \to \mp \infty$ if $\lambda \to \pm\infty$, we find that
 when $\gamma=0$, the graph of $p(\lambda,0)$ is an ``S'' shaped function passing through the origin.
 When $\gamma \geq 0$,  the graph is a shift up by $\frac{\gamma}{c}\delta_2q(u^*)$ to that of
  $p(\lambda,0)$  (See Fig. 4).

\begin{figure}[ht]
\setlength{\unitlength}{1mm} \centering \label{fig3}
\begin{picture}(60,50)
\put(30,0){\vector(0,1){50}} \put(0,17){\vector(1,0){70}}
\put(34,20){$\gamma<\gamma^*$} \put(32,13){\footnotesize$0$}
\put(57,2){$\gamma=\gamma^*$} \put(34,45){$\gamma>\gamma^*$} \put(68,13){$\lambda$}
\put(18,47){$p(\lambda,\gamma)$} \qbezier(1,25)(15,-5)(30,21)
\qbezier(30,21)(45,40)(55,1) \qbezier(1,33)(15,3)(30,29)
\qbezier(30,29)(45,48)(60,5) \qbezier(1,41)(15,11)(30,37)
\qbezier(30,37)(45,56)(65,8)
\end{picture}
\caption{The graph of $p(\lambda,\gamma)$ for $\gamma\geq 0$}
\end{figure}
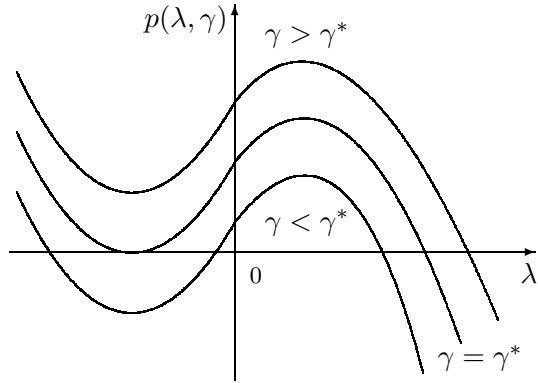

Therefore, if $0<\gamma$, the equilibrium point $E^*$ is hyperbolic, with a two dimensional ${W}^s_{loc}(E^*)$ and a one dimensional ${W}^u_{loc}(E^*)$.

From Fig. 4, it is easily seen that there is a threshold value
$\gamma^*=\gamma^*(A,B,\beta,c)$. If $0<\gamma< \gamma^*$, there are two distinct
negative real eigenvalues for $E^*$. If $\gamma=\gamma^*$, there is a repeated
negative real eigenvalue. If $\gamma>\gamma^*$, there is a complex conjugate
pair of eigenvalues with negative real part.

Hence, if $0<\gamma\leq \gamma^*$, $E^*$ has the two real negative eigenvalues.  Let $\mathbf{y}(s)$ be a solution  to \eqref{equ6} that is on
${W}^s_{loc}(E^*)$. Then $\mathbf{y}(s)$ approaches $E^*$ monotonically if $s$ is sufficiently large. If $\gamma>\gamma^*$,   $\mathbf{y}(s)$ approaches $E^*$ with damped oscillations as $s\to\infty$.

The negative root $\lambda_-$ of \eqref{derivative} is
$$\lambda_- =\frac{1}{3}\left(\omega_c-\sqrt{\omega_c^2+3\delta_1}\right)
$$
with $\omega_c=\frac{-\delta_1}{c}+c$.
There exists a unique $\gamma^*$ such that
$$p(\lambda_-, \gamma^*) = -(\lambda_-)^3+\omega_c(\lambda_-)^2
+\delta_1\lambda_- +\frac{\gamma^*}{c}\delta_2q(u^*)=0.
$$
Using $p'(\lambda_-,\gamma^*)=0$ to simplify $p(\lambda_-, \gamma^*)$ (long
division), the remainder is a first order polynomial in $\lambda_-$.
Some calculations show that
$$\gamma^*=\frac{2c}{27\delta_2q(u^*)}[(\omega_c^2+3\delta_1)\sqrt{\omega_c^2+3\delta_1}-(\omega_c^3+\frac{9}{2}\delta_1\omega_c)].$$
This completes the proof of (3) in Theorem \ref{th2}.

\end{proof}

\end{document}